\documentclass[manuscript]{aastex}
\usepackage{psfrag}
\usepackage{graphicx,shapepar}
\usepackage{lscape}
\usepackage{longtable}
\usepackage[title,titletoc]{appendix}
\usepackage{amsmath}

\newcommand{\vexp}{$V_{\textrm{\scriptsize exp}}$}
\newcommand{\kms}{\ensuremath{\mathrm{km\ s}^{-1}}}

\newcommand{\vsys}{\ensuremath{V_\mathrm{sys}}}
\newcommand{\grados}{$\,^\mathrm{o}\,$}

\newcommand\NIIlam{[N~II]\,6584\,\AA\@}

\newcommand\NII{[N~II]}
\slugcomment{Accepted for publication in the Astrophyiscal Journal}

\begin{document}

\title
{Internal Proper-Motions in the Eskimo Nebula}

\author
{Ma.~T. Garc\'{\i}a-D\'{\i}az\altaffilmark{1}, L. Guti\'errez\altaffilmark{1},
W. Steffen\altaffilmark{1}, J. A. L\'opez\altaffilmark{1} and J. Beckman\altaffilmark{2,3,4}}
\altaffiltext{1}{Instituto de Astronom\'{\i}a, Universidad Nacional Aut\'onoma de
  M\'exico. Km 103 Carretera Tijuana-Ensenada, 22860 Ensenada, B.C.,
  M\'exico}
\altaffiltext{2}{Instituto de Astrof\'{\i}sica de Canarias. La Laguna, Tenerife, Spain.}
\altaffiltext{3}{Consejo Superior de Investigaciones Cient\'{\i}ficas, Spain.}
\altaffiltext{4}{Departamento de Astrof\'isica. Universidad de La Laguna. Tenerife, Spain}
\email{tere,leonel,wsteffen,jal@astro.unam.mx, jeb@iac.es}

\begin{abstract}

We present measurements of internal proper motions at more than five
hundred positions of NGC~2392, the Eskimo Nebula, based on images
acquired with WFPC2 on board the Hubble Space Telescope at two epochs
separated by 7.695 years. Comparison of the two observations shows
clearly the expansion of the nebula. We measured the amplitude and
direction of the motion of local structures in the nebula by
determining their relative shift during that interval. In order to
assess the potential uncertainties in the determination of proper
motions in this object, and in general, the measurements were
performed using two different methods, used previously in the
literature. We compare the results from the two methods, and to
perform the scientific analysis of the results we choose one, the
cross-correlation method, as the more reliable.  We go on to perform a
``criss-cross'' mapping analysis on the proper motion vectors which
helps in the interpretation of the velocity pattern.  Combining our
results on the proper motions with radial velocity measurements
obtained from high resolution spectroscopic observations, and
employing an existing 3D model, we estimate the distance to the nebula
as 1300 pc.

\end{abstract}

\keywords{Planetary Nebulae: individual (NGC~2392) $-$ ISM:
  kinematics $-$ techniques: spectroscopic}

\maketitle
\section{Introduction}
\label{sec:introduction}

NGC~2392 ($\alpha$ = 07:29:10.76, $\delta$ = +20:54:42.47 [J2000.0])
is one of the most extensively studied high-ionization double-shell
planetary nebulae (see \citet{GDMT12}, hereafter Paper I, and
references therein) which is better known as the Eskimo nebula. The
Eskimo nebula shows a very complex structure: a main inner shell with
a filamentary shape surrounded by a ribbed structure (in Paper I
``caps''), an outer shell, bright cometary knots and collimated
high-velocity bipolar outflows.

Comprehensive spectroscopic kinematic studies in Paper I have shown
that the expansion velocity of the inner shell is \vexp$\approx
\pm$120~\kms. This kinematic study also revealed that the knots are
distributed in a disk very near the plane of the sky expanding at
velocities of $\approx$70~\kms.  \citet{ODell90} suggests that the
outer shell is an oblate spheroid and the inner shell can be well
described as a prolate spheroid oriented pole-on. In Paper I we found
that the inner shell is tilted by 9$^{\circ}$ with respect to the line
of sight with a position angle, P.A., of 25$^{\circ}$ and, to a first
approximation, has a width to length ratio of approximately 1.8.

The central star (CS) of the Eskimo has been studied by several
authors. An important discussion about the temperature of the CS is
given by \citet{Pottasch08}, in a paper about the abundances of the
nebula using spectral data obtained in the mid-infrared with the
Spitzer Space Telescope. The authors assume that [N~II] was formed
probably in the first dredge-up, that the abundance of carbon was
produced during the third dredge-up and that there is no evidence
about the existence of a second dredge-up. From this analysis the
authors found that the central star must have evolved from a
progenitor of 1.7~M$_{\odot}$. Several authors have calculated an
effective temperature, T$_{\mathrm {eff}}$, of the CS of around
40~000~K~--~45~000 K \citep{Mendez11, Pauldrach04,
  Kudritzki97}. However this temperature is not high enough to explain
high stages of ionization of some ions such as O~IV and Ne~V which
have been observed in the nebula \citep{Pottasch08,Natta80}.
\citet{Ciardullo99} observed a weak companion of the CS which is
undetected in V band.  In an attempt to model the companion star,
\citet{Danehkar12} used photoionization models and their result showed
that the companion star must have a T$_{\mathrm {eff}}${}~=~250 000 K,
which is much higher than that proposed by \citet{Pottasch08}. To date
there is no decisive evidence on the nature of the putative companion
star.

In order to understand the nature and origin of the Eskimo nebula, it
is crucial to know its distance, which at present is not very well
known. Several attempts have been made to find it using different
statistical methods. \citet{Maciel81} estimated a distance of 1.1~kpc
by using a mass-radius relation (Barker 1978). \citet{Hajian95}
measured the angular expansion of NGC~2392 at a radio frequency of
5~GHz with the VLA with 6 years between epochs, employing the Doppler
expansion velocity to calculate the distance to the nebula (assuming a
spherical shape). The authors did not detect an angular expansion of
the nebula, leading to a lower limit for the distance of
1.4~kpc. \citet{Stanghellini08} revised the calibration of the PN
distance scale from Cahn et al. (1992) using data for PNe in the
Magellanic Cloud. This statistical method is based on a calibration of
the relation between the ionized mass (assuming that all nebulae have
the same ionized mass) of PNe and the optical thickness parameter. In
this way, the authors obtained a distance for NGC~2392 of 1.6~kpc with
an uncertainty of 130 pc. \citet{Pottasch11} reported 1.8~kpc for the
distance, inferred from the core mass, the chemical nebular
abundances, and the luminosity.

Another method for calculating distances is using proper motion
measurements. Sample of a previous application of proper motion
methods is found in Artigau et al (2011), who used cross-correlation
methods to measure the proper motions of the knots and arcs of
Eta Carinae using data from the Near-Infrared Coronagraphic Imager and
NACO. Using the same method (cross-correlation), Ueta et al. (2006)
calculated proper motions of the dust shell structure in the Egg
Nebula based on the archived two-epoch data at 2 $\mu$m taken with the
Hubble Space Telescope ({\it HST}, with a 5.5 yr interval). From these
data the authors determined the distance to the Egg nebula.

Szyszka et al (2011) used a different method on two epochs of {\it
  HST} imaging, separated by 9.43 yr in order to measure the expansion
proper motions for NGC~6302's bipolar lobes, calculating the $\chi^2$
of the difference image.

Li et al. (2002) observed the PN BD+30-3639 using the WFPC2 camera on
board the {\it HST}.  The data were obtained from two different epochs,
separated by 5.663 yr.  They used the $\chi^2$ method to measure the
radial expansion of this PN.  To derive the distance of the nebula,
these authors combined the angular expansion with radial expansion
velocities taken from the {\it HST STIS} Echelle spectrograph.

A recent paper about proper motions was published by O'Dell et
al. (2013), where tangential velocities were calculated for NGC~6720
using the least-squares ($\chi^2$) method.

These studies all present the results of either only cross-correlation
or the $\chi^2$ method for their determination of internal proper
motion patterns. Up to now it is, however, unclear how these methods
compare and how consistent are the results when applied to the same
data set. In order to asses better the reliability of proper motion
measurements in general and for this object in particular by these
methods, we apply both of them (cross-correlation and $\chi^2$) on the
same data sets.

In this paper, we report proper motion measurements for a large number
of structures and arcs in the Eskimo nebula, using two \NIIlam\,
images from the {\it HST} archive which were observed with a time
interval of 7.695~yr. The results, along with the radial velocity
measurements taken from long-slit, high-resolution, spectroscopic
observations (Paper I) allow us to determine the distance.

The structure of the paper is as follows.  In \S\ref{sec:observations}
we describe the observations and the data-reduction steps. In
\S\ref{sec:measurements} we describe in more detail the methods for
finding the measurements of the proper motions in the nebula. In
\S\ref{sec:differences} we discuss the differences between the
methods. In \S\ref{sec:radial} we discuss the radial velocities. In
\S\ref{sec:distance} we explain the method for calculating the
distance. In \S\ref{sec:criss-cross} we perform a criss-cross mapping
analysis. Finally, we present in \S\ref{sec:discussion} and
\S\ref{sec:conclusion} a discussion of the results and our
conclusions.

\section{Observations and Data Reduction}
\label{sec:observations}

\subsection{High-resolution spectroscopy}

For the present work, we used the data from Paper I. They are
high-resolution spectroscopic observations of the Eskimo nebula
obtained at the Observatorio Astron\'omico Nacional San Pedro
M\'artir (OAN-SPM), Baja California, M\'exico, during the nights of
January 7--10, 2002. The data were obtained with the Manchester
Echelle Spectrometer \citep{Meaburn03} attached to the 2.1~m telescope
in its f/7.5 configuration.  For all positions we used a 90 \AA\,
bandwidth filter to isolate the 87th order containing the H$\alpha$
and [N~II] nebular emission lines. For the majority of the exposures,
we used a 70$\mu$m slit, while for some exposures a 150$\mu$m
slit was applied. The slit was oriented north-south, and exposures
were taken at a series of 18 different parallel pointings and three
pointings at other position angles, P.A.~=~70$^{\circ}$ (two
positions) and P.A.~=~110$^{\circ}$ (one position). The positions and
orientations of the slit are shown in Figure~1 of Paper I.  All
spectra were acquired using exposure times of 1800~seconds. Full details of
the observations and data reduction process are given in Paper I.

We convert heliocentric radial velocities to velocities relative to
the Eskimo nebula by taking the systemic velocity, \vsys~=~70.5~\kms,
calculated in Paper I. The spectroscopic data are summarized in
Table~1.

\begin{table*}[!h]
\centering
\caption{Log of spectroscopy observations from OAN-SPM}
\vspace{0.2cm}
\begin{tabular}{cccccc} \hline\hline
Position$^\dagger$ & Data &  Slit  & PA\\
  & yyyy/mm/dd  & $\mu$m & \grados\\ \hline
a& 2002-01-09 & 150 & 0\\
b -- c& 2002-01-08  & 70 & 0\\
d& 2002-01-09  & 70  & 0\\
e& 2002-01-08  & 70  & 0\\
f& 2002-01-09  & 70  & 0 \\
g& 2002-01-07  & 70  & 0\\
h -- n & 2002-01-08 & 70 & 0\\
o -- q& 2002-01-09 & 70  & 0\\
r& 2002-01-09  & 150 & 0\\
s& 2002-01-10  & 70  & 110\\
t& 2002-01-10  & 70  & 70\\
u& 2002-01-10  & 70  & 70\\
\hline
\end{tabular}
\begin{tabular}{l}
$^\dagger${The Positions of the slit are shown in Figure~1}
of Paper I
\end{tabular}
\label{tab:log-spec}
\end{table*}

\subsection{Hubble Space Telescope Data}

The data used in this study to measure the proper motions of the
Eskimo nebula consist of images which were retrieved from the {\it HST}
archive. The observations were made in two separate observing
runs: on January 1, 2000 (hereafter {\it Epoch-1}),
 as part of program 8499, with Andrew S. Fruchter as PI. During this
 run, three images were obtained with the Wide Field Planetary Camara 2
 (WFPC2), with exposure time 350~s, using (among others)
 the F658N filter. The second-epoch data
 were obtained on September 21, 2007 (hereafter {\it Epoch-2}),
from the science program 11122, with Bruce Balick, as PI, who employed
the same configuration as in program 8499 to acquire three images with
the F658N filter, using 400~s exposure time for all of them. For
  all of the images we used the drizzled images (processed with
  Astrodrizzle).  Table~2 summarizes the imaging observations.

\begin{table}[!h]
\centering
\caption{Log of photometry observations from {\it HST}}
\vspace{0.2cm}
\begin{tabular}{cccccc} \hline\hline
 Date   &Exposure &Dataset \\
 yyyy/mm/dd &  sec & \\ \hline
2000-01-11  & 350 &  u60v010er\_drz \\
2000-01-11  & 350 &  u60v010fr\_drz \\
2000-01-11  & 350 &  u60v010gr\_drz \\
2007-09-21  & 400 &  ua010505m\_drz \\
2007-09-21  & 400 &  ua010506m\_drz \\
2007-09-21  & 400 &  ua010507m\_drz \\
\hline
\end{tabular}
\label{tab:log-photo}
\end{table}

The time interval between observations is 7.695~yr with an angular
resolution of $\approx$ 0$\farcs$1. The mosaicing, geometric
correction, and astrometry correction were done in the HST pipeline
using Astrodrizzle.  The mosaics were median combined.  In order to
compare the two combined mosaics, we regridded them using the SWARP
utility \citep{Bertin02}, matching the X-Y directions with RA-DEC.
  The correction of the very small residual relative displacements
  between the two mosaics were performed recursively by analyzing
  visually the arithmetic difference, using as reference the general
  structure that remains fixed. The shifting was carried out in
  0.1-pix steps.

We used the central star as reference. The positions of the points
where we calculated the proper motions in the nebula are given as
offsets in RA and DEC referred to this point.

\begin{figure*}[!t]
\centering
  \includegraphics[width=1.\textwidth]{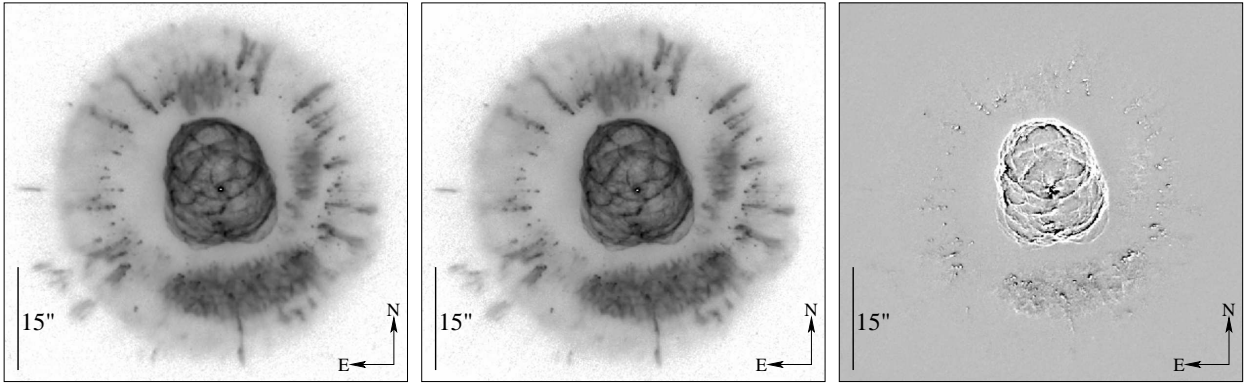}
  \caption{Left: {\it Epoch-1} image \NII{} WFPC2-{\it HST}. Center:
    {\it Epoch-2} image \NII{} WFPC2-{\it HST}. Right: Difference of
    the two images taken 7.695 yr apart (2007-2000). In the difference
    image, the central bubble with proper motions shows up as
    ``negative/positive'' double ridge structures. }
  \label{fig:Figure1}
\end{figure*}

The resulting pictures are shown in Figure~\ref{fig:Figure1}, which shows in the left
and the center panels the two final images for each epoch (2000.03 and
2007.72 respectively) centered taking as reference the center of the
Eskimo nebula, and the difference image from the two epochs in the
right panel.  In general, we can see that the arcs of the inner
bubble have outwards motion from the star, whereas the knots of the
outer shell have more modest movements than the arcs of the
central bubble.

\section{Measurement Technique}
\label{sec:measurements}

In order to measure the proper motions in the nebula, we determined
the shift of $>$ 500 regions, defined using well-defined knots or
arcs. We used two methods based on the assumption that the proper
motion of any local structure in a nebula due to expansion can be
measured by determining the translational shift of the structure.

{\bf First method:} This method employs cross-correlation, sometimes
called \emph{cross-covariance}. It is a measure of the similarity of
two signals, in the one-dimensional case, or of two images, in the
two-dimensional case. It is used mainly to look for patterns in an
unknown signal comparing it with a known one.

The definition of the cross-correlation function $C_{fg}(i,j)$
normalized in a digital discrete image is given by:

\begin{equation}
C_{fg}(i,j) =
\frac{\sum\limits_{x=1}^M\sum\limits_{y=1}^N\Big[f(x,y)-\bar{f}\Big]\Big[g(x-i,y-j)-\bar{g}\Big]}{\sigma_f
  \sigma_g} \label{eq:n1}
\end{equation}

\noindent
where $f(x,y)$ and $g(x,y)$ are functions representing the pixel
values of the two images, {\it i} and {\it j} are the relative
displacements of one image ({\it Epoch-2}) with respect to the other
({\it Epoch-1}), $\bar{f}$ and $\bar{g}$ are the average values of
functions $f$ and $g$, respectively, $M$ and $N$ are the dimensions of
the subimage used (in our case, instead of analyzing the entire image
we analyzed sections of the image, and we refer to those as
\emph{subimages}), and $\sigma_f$ and $\sigma_g$ are their standard
deviations. Then, $C_{fg}(i,j)$ varies between $-1$ and $1$.

In this case, to calculate the velocity of a given point, we defined a
\emph{box} of size $N \times N$ pixels around that point, where $N$ is
chosen according to the characteristics of the neighborhood of the
point.  Generally, $N$ = 15 was enough to measure the displacements of
the points.

We denote as $p$ the center of the box. This $N \times N$ pixels box
is correlated with a box of the same size centered at $p + \delta$ in
the other image. We then move $\delta$ so that $-\delta_{max} \leq
\delta_{x} \leq \delta_{max}$ and $-\delta_{max} \leq \delta_{y} \leq
\delta_{max}$, obtaining a value for the correlation at each point, to
form the function $C(p,\delta)$. The point with coordinates
$(\delta_{x},\delta_{y}$) where the function is maximum determines the
displacement of the structure under study.

Calculating the cross-correlation for different displacements,
$\delta$, in one of the images, we can build the function
$C(p,\delta)$ = $C_{fg}(\delta_{x},\delta_{y})$.
For the calculations, we used the task XREGISTER implemented in IRAF, where
the cross-correlation function is computed in the following  manner:

\begin{equation}
C_{fg}(\delta_{x},\delta_{y}) =
\frac{\sum\limits_{x=1}^M\sum\limits_{y=1}^N\Big[f(x+\delta_{x},y+\delta_{y})-\bar{f}\Big]\Big[g(x,y)-\bar{g}\Big]}{\sigma_f
  \sigma_g} \label{eq:n1}
\end{equation}

\noindent where

\begin{equation}
\sigma_f = \Big[\sum\limits_{x=1}^M\sum\limits_{y=1}^N\Big[f(x+\delta_{x},y+\delta_{y})-\bar{f}\Big]^2\Big]^{\frac{1}{2}}
\end{equation}

\begin{equation}
\sigma_g = \Big[\sum\limits_{x=1}^M\sum\limits_{y=1}^N\Big[g(x,y)-\bar{g}\Big]^2\Big]^{\frac{1}{2}}
\end{equation}

\begin{equation}
\bar{f} = \frac{\sum\limits_{x=1}^M\sum\limits_{y=1}^N\Big[f(x+\delta_{x},y+\delta_{y})\Big]}{M \times N}
\end{equation}

\begin{equation}
\bar{g} = \frac{\sum\limits_{x=1}^M\sum\limits_{y=1}^N\Big[g(x,y)\Big]}{M \times N}
\end{equation}

\begin{figure*}[!t]
    \centering
    \includegraphics[width=1\textwidth]{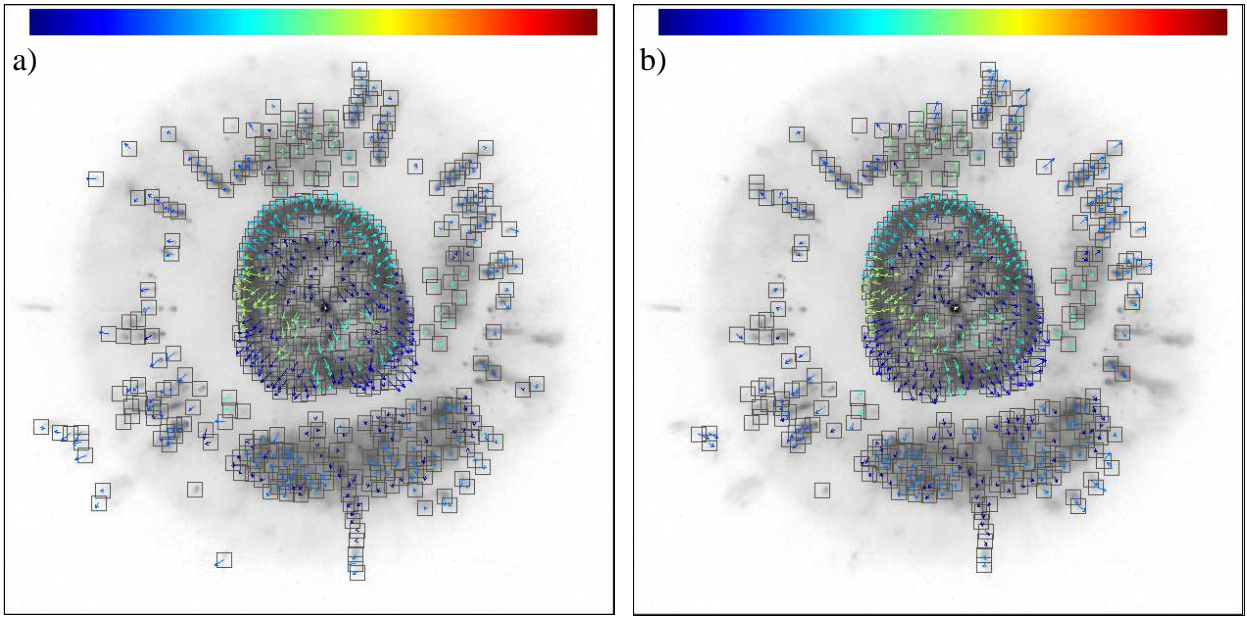}
  \caption{{\it Left panel:} Map of proper motions obtained with the
    cross-correlation method. {\it Right panel:} Map of proper motions
      obtained with the $\chi^2$ method}
  \label{fig:Figure2}
\end{figure*}

Repeating the calculation  in all the 536 regions marked on the {\it
  Epoch-1} image, we obtained the displacement map shown in Figure~\ref{fig:Figure2}
(left panel), representing the internal motion of the nebula in the
plane of the sky.

{\bf Second method:} This method is based on the minimization of
chi-squared ($\chi^2$). We used the same 536 different local regions
identified above.

\begin{figure*}[!h]
    \centering
    \includegraphics[width=1\textwidth]{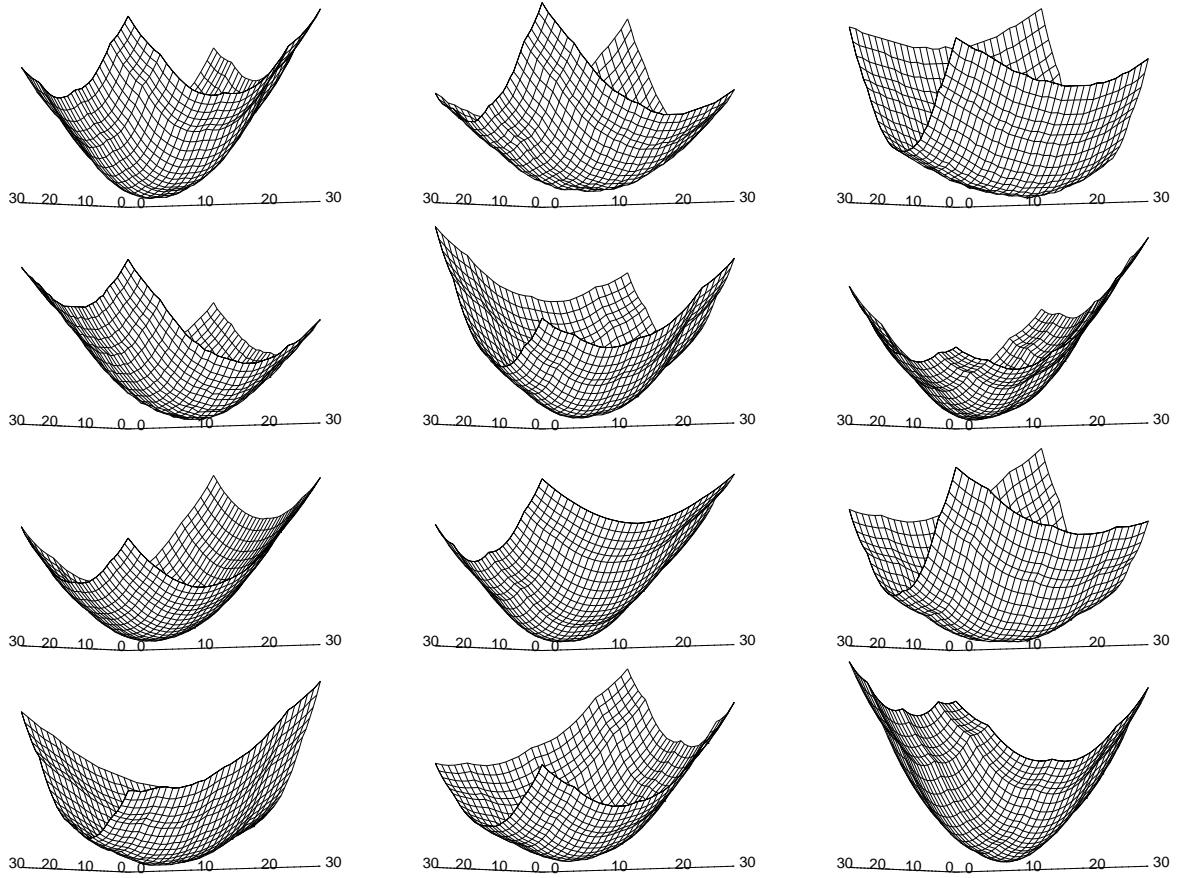}
  \caption{Some of the 3-D plots of chi-squared as a function of the
    displacements in X and Y. The scale of the Z axis is in arbitrary
    units. In this figures we can see the minimum very well defined.}
  \label{fig:Figure3}
\end{figure*}

These structures should be located at slightly shifted positions in
the {\it Epoch-2} images due to proper motion of the nebula. To
quantify the displacements, we defined square image sections centered
on each of these local structures in the {\it Epoch-1} image. The size
of these image sections has to be large enough so that the segmented
structures can be uniquely identified.  To analyze each section, we
produced a custom IRAF task to shift the section of one epoch respect
to the corresponding section of the other epoch, to calculate the
difference and to estimate the chi squared ($\chi^2$) on the
difference. The script repeats this process spanning from $-{\Delta}x$
to $+{\Delta}x$, and from $-{\Delta}y$ to $+{\Delta}y$, where
${\Delta}x$ and ${\Delta}y$ were programable, in increments of 0.1
pixel. Visually we determined that the displacements are generally
smaller than 1 pixel, so we carried out several trials putting
${\Delta}x = {\Delta}y$ = 2 or 3 pixels. In
Figure~\ref{fig:Figure3} we show a 3D plot of $\chi^2$ for 12 example
cases, where we can see that the surface shows a minimum value of
$\chi^2$ in these plots, which corresponds to the displacement for
which the region of the second epoch is most similar to the same
region on the first epoch. Then measuring the position of this minimum
gives the magnitud of the displacement vector, because the x,y
coordinates of this point are the relative positions given in
0.1-pixel steps with respect to the center of the region on the {\it
  Epoch-1} image.

\begin{figure*}[!h]
\centering
    \includegraphics[width=1\textwidth]{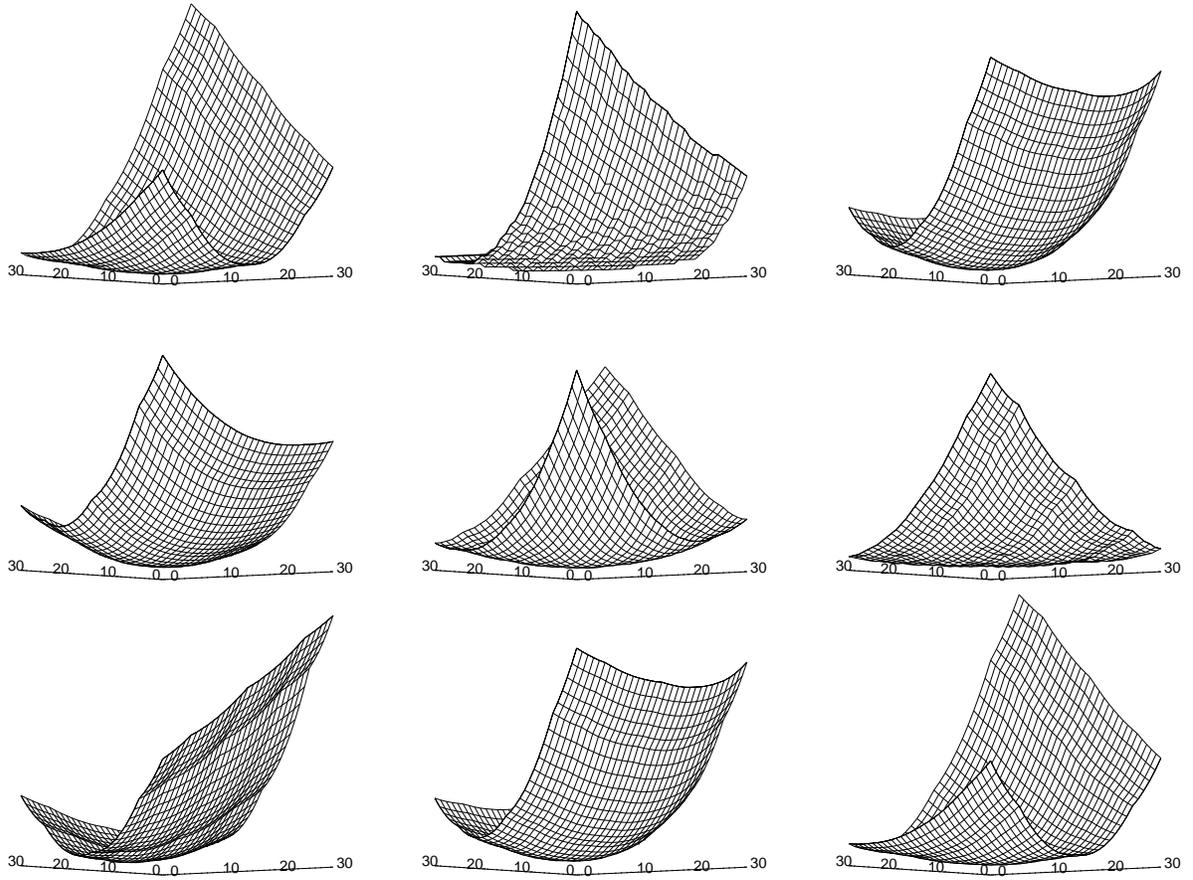}
  \caption{Same as Fig.~4, but can be noticed that in this case the
    function around the minimum is more difficult to be defined.  In
    several of these cases the calculated velocity vector was different than the
    movement observed using a visual approach.}
  \label{fig:Figure4}
\end{figure*}

In general, the map obtained with this method (see
Figure~\ref{fig:Figure2}, right panel) is a good representation of the
movements in the nebula.  However, we found several cases where the
minimum value is not so well defined as we can see in
Figure~\ref{fig:Figure4}. Generally in these cases the velocity vector
was different from the movements observed in a visual approach. The
reason for these discrepances could be the uncertainties in the
determination of the displacements, given by the center of the surface
of the $\chi^2$ plot.

\subsection{Differences between the two methods to calculate the proper motions.}
\label{sec:differences}

We measured the proper motions of more than 500 regions of NGC~2392 by
two methods: one using cross-correlation and another using the
minimization of $\chi^2$. There are several points where the magnitude
of the displacement obtained by minimizing the $\chi^2$ is larger than
a full pixel, but visual inspection shows that the movements are
smaller than that. We believe that these values, obtained by
minimizing $\chi^2$, are generally oversized in the process of
searching the minimum value, particularly in regions where the shape
of the surface defined by the value of $\chi^2$ as a function of the
displacements in X and Y is not as sharp and well behaved (see
Figure~\ref{fig:Figure4}) as those in Figure~\ref{fig:Figure3}.

\begin{figure*}[!h]
    \centering
    \includegraphics[width=1\textwidth]{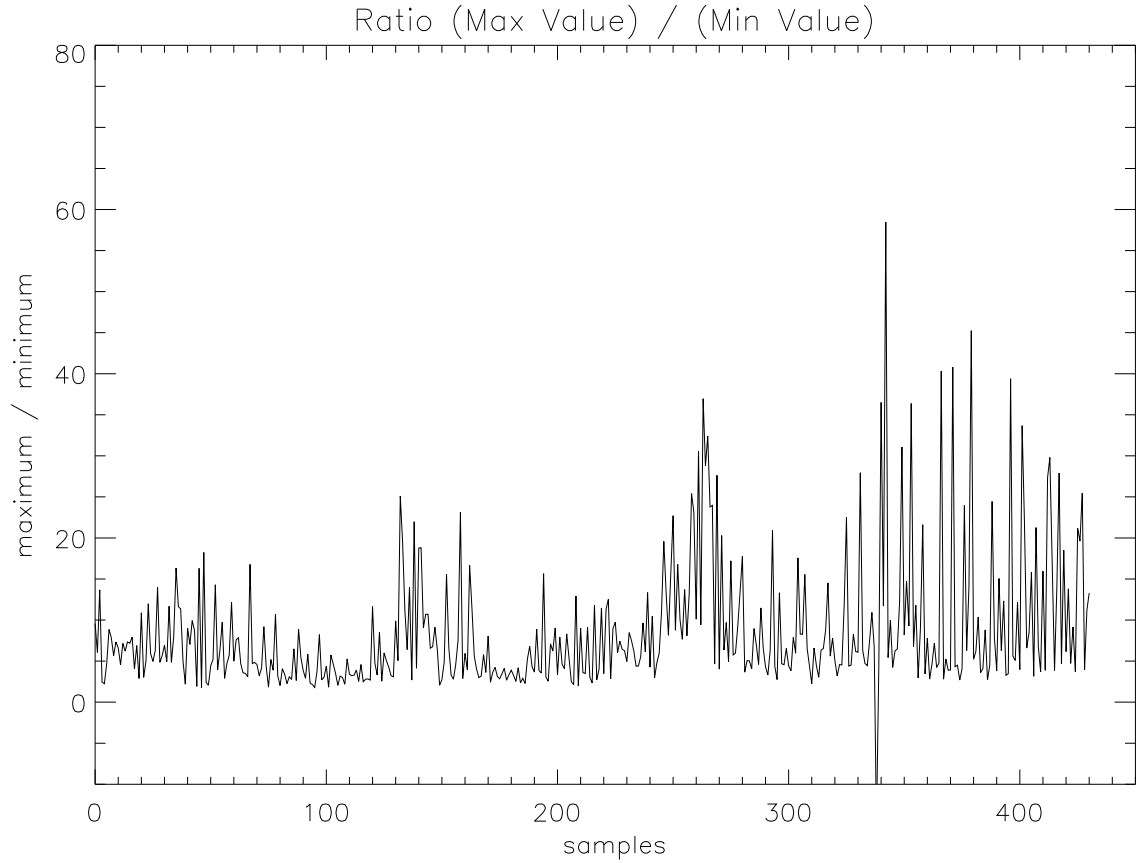}
  \caption{A plot of the ratio {\it maximum/minimum} value for the
    pixels for each of the 536 regions selected in the $\chi^2$
    method, having discarded those for which the magnitude of the
    displacement is larger than a full pixel. We can see that in some
    cases this ratio is very large or negative. The larger and the minimum
    values correspond in this case to regions near of the central star, where
    we found saturated pixels.}
  \label{fig:Figure5}
\end{figure*}

There are a couple of other points whose associated
regions have very high ratio between the maximum and minimum
values, or where this ratio is negative. In Figure~\ref{fig:Figure5}
we show a plot of the ratio maximum/minimum of the pixel values in
each subregion considered for each point in the $\chi^2$ method, taken
as a measure of the contrast. We see those cases in Figure~\ref{fig:Figure5}.

Discarding those points we see
that in the overall, the discrepancy between these two methods is really
small. However, we preferred the values obtained with the
cross-correlation method since we did not need to discard any values
with that method. In addition, comparing the two images using a direct
visual approach, we can see that the values obtained from the
cross-correlation method are more representative of the motion in the
nebula, since the global proper motions are more readily apparent in
the cross-correlation method.

\section{Radial Velocities}
\label{sec:radial}

\begin{figure*}[!t]
    \centering
    \includegraphics[width=1\textwidth]{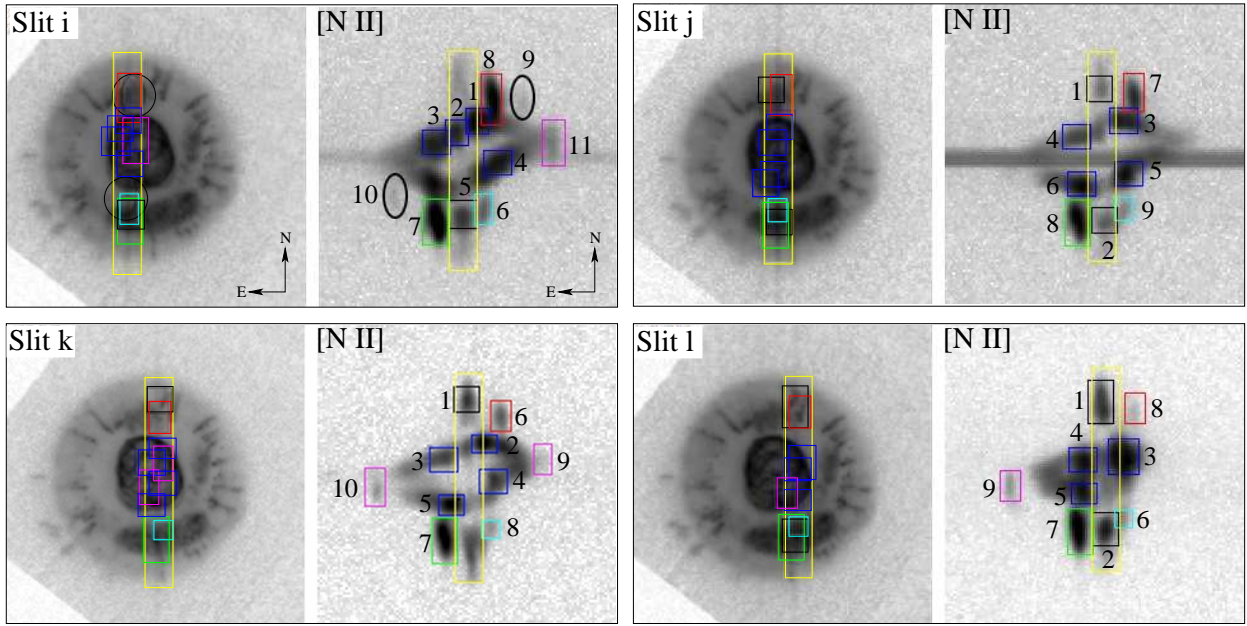}
  \caption{Example of regions on slits {\it i — l} where we measured the radial velocity. Details about these observations are in Paper I.}
  \label{fig:Figure6}
\end{figure*}

The radial velocity of each local big structure or region (see
Figure~\ref{fig:Figure1}) was calculated using the \NIIlam{} line profile for 20
individual slit position originally used for Paper I. We identify each
slit position over an {\it HST} \NII{} image relating each component
of the emission profile with the corresponding region of the {\it HST}
image. In the position-velocity (P--V) arrays shown in Figure~2 of
Paper I, we found several distinct regions distributed along the slit,
Figure~\ref{fig:Figure6} shows an example of the distribution of the regions in slits
{\it i} -- {\it l}.

These regions include the jets but we did not use them because they
are not visible in the {\it HST} images. We calculated the
heliocentric velocity for each region by fitting a gaussian to the 1-D
profile. The radial velocities are then calculated subtracting the
systemic velocity. The errors in the determination of the velocities
are given by the $\sigma$ in each gaussian fitted, according to the
relation $FWHM = 2\sqrt{2\ln(2)}\sigma$. The median of those values of
$\sigma$ is 10.5 \kms.

In some cases, we had problems with the identification of the
velocities of some of the regions, because we found two or three
different velocities in the same position of the {\it HST} image. In
those cases, we took the velocity of the brightest node in the P--V
array.

The radial velocities with respect to the central star are listed in
Table~3.

\section{Distance}
\label{sec:distance}

\begin{figure*}[!t]
    \centering
    \includegraphics[width=1\textwidth]{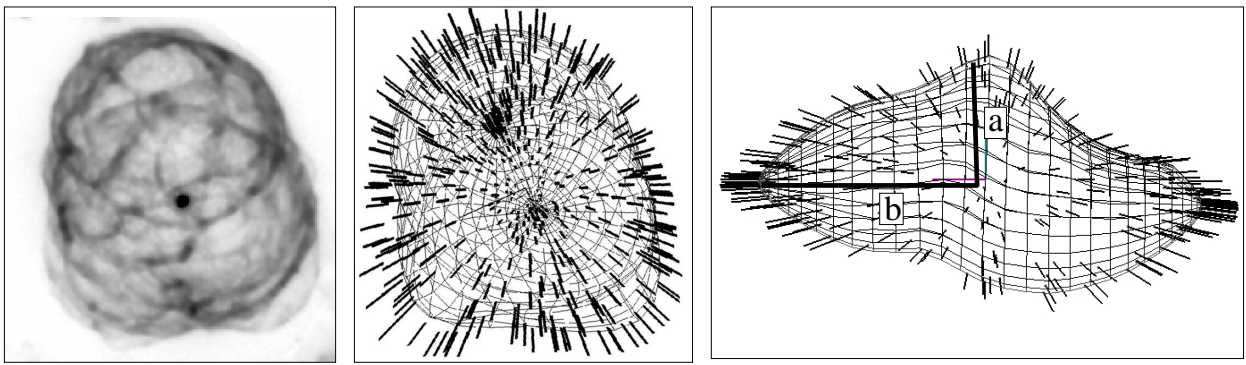}
  \caption{The inner bubble of Eskimo nebula and the velocity vectors according to the model (Paper I).}
  \label{fig:Figure7}
\end{figure*}

We calculate the distance to the Eskimo nebula using the proper motion
vectors calculated from the cross-correlation method and taking only
the vectors of the inner shell, given that we know its
geometry. Statistically, the components of proper motions are $\left<
pm_x \right>$ $\approx$ $\left< pm_y \right>$.  If we consider a
spherical distribution, the root mean square (RMS) of the radial
velocity would be: $\left< rv \right>$ $\approx$ $\left< pm_x \right>$
$\approx$ $\left< pm_y \right>$. Assuming the shape of the inner shell
as modeled in Paper I (see Figure~\ref{fig:Figure7}) where the ratio between the major
axis (b) and minor axis (a) is b/a = 1.8, therefore,

\begin{equation}
\left<rv\right> = (b/a) \left<pm_x\right> = (b/a) \left<pm_y\right>,
\end{equation}

or, in terms of the RMS of the proper motions ($\left<pm\right> =
\sqrt{pm_x^2 + pm_y^2}$)

\begin{equation}
\left<rv\right> = (b/a) \frac{\left<pm\right>}{\sqrt{2}} \label{eq.a}
\end{equation}

If we express the time interval between the two epochs of observation
by $\delta t$ (in years), the rms of the measured transverse
displacements (on the images) by $\left<\delta r\right>$ (in arcsec),
and the distance $D$ (in parsecs) to the Eskimo nebula, the rms of
proper motions can be expressed by (see Appendix~\ref{ap:a},
eq.~\ref{eq:n6}),

\begin{equation}
          \left<pm\right> =
          4.74 \frac{D~\left<{\delta}r\right>}{{\delta}t} \label{eq:n2}
\end{equation}

Using the eq.~\ref{eq.a}, we can compute D by,

\begin{equation}
          D = 0.211~\frac{{\delta}t}{\left<{\delta}r\right>}\,
          \frac{\sqrt{2} \left<rv\right>}{(b/a)} \label{eq:n3}
\end{equation}

\noindent
and from our measurements we calculate a distance of 1300 pc. This
value is similar to other published values: ~1.4~kpc from
\citet{Hajian95}, 1.6~kpc from \citet{Stanghellini08} and 1.8~kpc from
\citet{Pottasch11}.

Applying our newly computed distance, Table~3 lists the proper motion
values converted from angular displacements to velocity in km
 s$^{-1}$, where we have used the scale factor to convert arcsec into
\kms\, from Eq.~\ref{eq:n2} (801.3 km sec$^{-1}$ arcsec$^{-1}$
calculated in Appendix~\ref{ap:a}, Eq~\ref{eq:n7}).



\section{Criss-cross mapping}
\label{sec:criss-cross}

Criss-cross mapping was recently developed by Steffen \& Koning (2011)
as an analysis tool to identify patterns in proper motion
measurements, in particular systematic deviations from homologous
expansion. The basic idea is to find regions where the projected
velocity vectors converge or diverge. For the mapping, the vectors are
extended over the full image range independently of the direction.
All lines have a fixed brightness value (e.g. 1) and are then added
together in an image. The resulting image, which may be convolved with
a gaussian smoothing function, will show enhanced values where the
line crossing-points cluster, thereby revealing regions on which the
motion converges or from where it diverges.

As can be appreciated by close inspection of Figure~\ref{fig:Figure2}, the proper
motion pattern obtained by applying the cross-correlation and the
$\chi^2$ methods are somewhat different from each other and clearly not
consistent with the radial pattern expected from homologous expansion, which
produces a central point-like concentration (Steffen \& Koning, 2011). In
order to assess whether the deviations from homologous expansion are
purely intrinsic to the methods or might contain information about
deviations of the 3D velocity field from homologous expansion, we
perform a criss-cross mapping analysis on the data and on a correction
to the model of homologous expansion. The model is based on the
following hypothesis.

Since the inner bubble of the Eskimo nebula contains hot X-ray
emitting gas (Guerrero et al. 2005; Ruiz et al. 2013) it may well be
that its expansion is dominated by thermal pressure. Therefore,
instead of expanding homologously, the direction of the velocity field
might be perpendicular at every surface point of the bubble. In
Figure~7 (left panel) we therefore show the expected proper motion pattern for an
axi-symmetric ellipsoid with axis ratio of 1.8 at an inclination angle
of 20$^\circ$, velocity vectors that are perpendicular to the local
surface.

The predicted proper motion pattern is not unlike the observed pattern
in that it contains strongly deviating vector directions and
magnitudes very close to each other.  This is because in this type of
models the vectors from the front and back portions of the nebula may
in fact have different magnitudes and directions at the same projected
positions. In a homologous expansion the projections of all vectors
onto the sky are radial, no matter what the orientation or whether
they come from the front or back. For a more detailed qualitative
analysis we apply the criss-cross mapping technique to the
observations and this model (Figure~8).

\begin{figure*}[!t]
    \centering
    \includegraphics[width=1\textwidth]{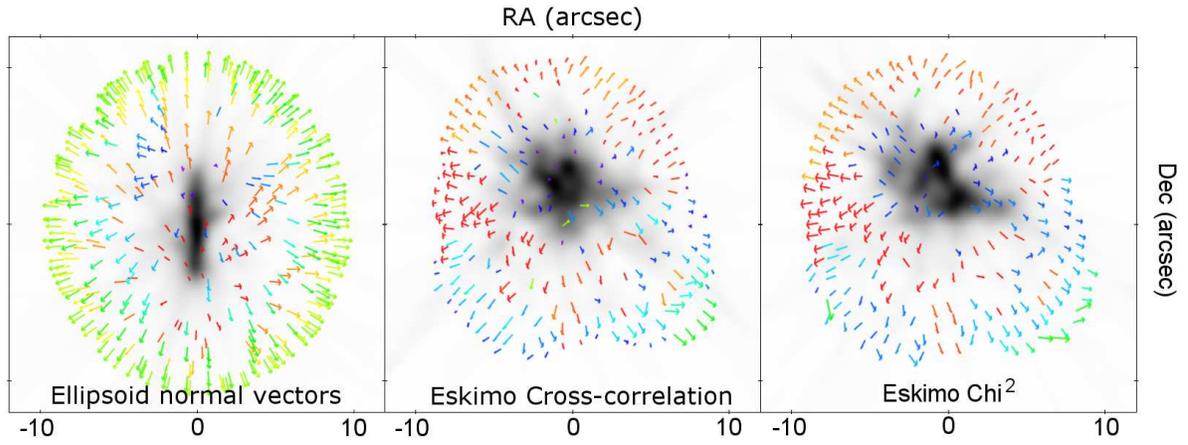}
  \caption{The criss-cross maps for the ellipsoidal model (left panel) and the measured values for the cross-correlation and $\chi^2$ methods in the middle and on the
    right, respectively.}
  \label{fig:Figure8}
\end{figure*}

 The criss-cross map for the non-homologous ellipsoidal model shows a
 very characteristic pattern of an approximately straight line with
 uniform brightness along the direction of inclination
 (Figure~\ref{fig:Figure8}, left). The criss-cross maps for the
 observations (Figure~\ref{fig:Figure8}, middle and right) show a more
 complex structure centered to the north of the central star. The
 overall structure is, however, much less elongated in the
   north-south direction compared to the ellipsoidal model and are
   more consistent with an off-center noisy point structure. The
   criss-cross mapping is therefore not consistent with an expansion
   perpendicular to the nebular surface but rather with radial
   expansion. The center of expansion is however off-center from the
   central star.

\section{Discussion}
\label{sec:discussion}

In this article we have produced proper motion velocity maps for the
Eskimo nebula, using images from the {\it HST} taken at two different
epochs separated by 7.695 years. One map was generated using a
method based on $\chi^2$ minimization and the other was obtained by
calculating the cross-correlation. Both methods used the same ( $>$
500) subregions of the images. We find that the cross-correlation
method provides a slightly more continuous pattern of proper motion
vectors than the $\chi^2$ method.

The X-ray emission, constrained by the outline of the inner bubble,
suggests the possibility that the expansion of the bubble is dominated
by the thermal pressure of the hot gas rather than by the inertia of
the bubble, resulting in an expansion perpendicular to its surface,
rather than radially outwards.  A comparison with a simple ellipsoidal
model with velocity vectors perpendicular to the surface shows that
such a model indeed reproduces the overall local deviations of
neighboring proper motion vectors.

Criss-cross mapping analysis is incompatible with the pressure
  driven overall ellipsoidal model and with the velocity perpendicular
  to the surface.

We also present the radial velocities calculated at the different
points where we measured the proper motions.  From those we infer a
distance to the nebula of 1300~pc, a value well within the range of
the published values (1.1 -- 1.8 kpc) for the distance to this nebula.

\section{Conclusions}
\label{sec:conclusion}

A key result of this study is that the application of two different
methods for the determination of internal proper motion in the Eskimo
Nebula (NGC~2392) based on the same observational data has shown that
the results are quite similar with minor deviations between the two
methods.

Criss-cross mapping of the proper motion vector field yields no evidence
in favor of an expansion perpendicular to the inner bubble, which in
turn is an indication that the hot gas inside the bubble is not driving
the expansion, rather than inertia of the dense shell of NGC~2392.

Last, but not least, based on our data we determined a distance to the
Eskimo Nebula of approximately 1.29~kiloparsec.

\section*{Acknowledgments}

This research has benefited from financial support from DGAPA-UNAM
through grants IB100613-RR160613 and PAPIIT IN101014, as well as from
Conacyt through grant 167236.  We acknowledge the excellent support of
the technical personnel at the OAN-SPM.  This work is partially based
on observations made with the NASA/ESA {\it Hubble Space Telescope},
obtained from the data archive at the Space Telescope Science
Institute. STScI is operated by the Association of Universities for
Research in Astronomy Inc., under NASA contract NAS5-26555. We thank
the anonymous referee for the constructive comments that improved the
presentation of this work.


\begin{appendices}
\section{}

\label{ap:a}

Using our proper motion measurements, we are able to calculate the
distance to the Eskimo nebula. Considering any distance subtended by
an angle $\delta r$ (in this case, $\delta r$ is the measured
transverse displacements in arcsec), as s. Then the distance, $D$
to the Eskimo nebula is calculated by
\begin{equation}
\begin{aligned}
s(\mathrm{km}) = D(\mathrm{km}) \delta r(\mathrm{rad}) =
3.0856\times10^{13}(\mathrm{km\, pc}^{-1})\, \\ \times
D(\mathrm{\mathrm{pc}})\,\frac{\delta r(\mathrm{arcsec})}{206264.8
  (\mathrm{arcsec}/\mathrm{rad})}. \label{eq:n4}
\end{aligned}
\end{equation}

Expressing the time interval between observations by $\delta t$, the
RMS of the measured proper motions is given by
\begin{equation}
\begin{aligned}
\left<pm(\kms)\right>\, =
\frac{S(\mathrm{km})}{\delta t(\mathrm{s})} =\\
\frac{3.0856\times10^{13}\, D(\mathrm{pc})\, \delta r(\mathrm{arcsec})\,
  /206264.8}{\delta
  t(\mathrm{yr})\times365.24\times24\times3600} \label{eq:n5}
\end{aligned}
\end{equation}

\begin{equation}
\left<pm(\kms)\right> = 4.7405\, \frac{D(\mathrm{pc})\, \delta
  r(\mathrm{arcsec})}{\delta t (\mathrm{yr})}. \label{eq:n6}
\end{equation}

Statistically, $\left< pm_x \right>$ $\approx$ $\left< pm_y \right>$,
so with a spherical distribution $\left< rv \right>$ $\approx$ $\left<
pm_x \right>$ $\approx$ $\left< pm_y \right>$.  However, assuming the
shape of the inner shell as modeled in Paper I (where the major and
minor axes are b and a), we can consider that, statistically, the RMS
of the proper motions are b/a times the RMS of the radial velocities,
$\left<rv\right>$, i. e.
\begin{equation}
\left<rv\right> = b/a \left<pm_x\right> = b/a \left<pm_y\right>,
\end{equation}
\noindent or
\begin{equation}
\left<rv\right> = b/a \frac{\left<pm\right>}{\sqrt{2}}.
\end{equation}

Then the distance to the nebula is
\begin{equation}
D(\mathrm{pc}) = 0.211 \frac{\delta t(\mathrm{yr})}{\delta
  r(\mathrm{arcsec})} \frac{\sqrt{2} \left<rv(\kms)\right>}{(b/a)}.
\end{equation}

\noindent
taking b/a = 1.84 which was derived from the 3D model (Paper I), we
final $D$ = 1300 pc. If we consider $\delta t$ = 8.805 yr, we are
able to calculate a scale factor to convert arcsec into \kms\, from
Eq.~\ref{eq:n6}

\begin{equation}
\left<pm(\kms)\right> = 801.3\, \delta r(\mathrm{arcsec})\label{eq:n7}
\end{equation}

\end{appendices}

\newpage

\vfill\eject
\begin{table*}
  \caption{Columns 1 and 2 give the position of each region, referred
    to the central star. Columns 3 and 4 give the proper motion in
    pixels of each knot. Columns 5 and 6 contain the proper motions in
    \kms. Column 7 gives de magnitud of the proper motion vector in
    \kms. Column 8 gives the radial velocity in \kms\, calculated from
    high resolution spectra (the full table is in the online version
    of this paper).}
  \label{T:3}
~\\
    \begin{tabular}{cccccccc}
\hline \hline
(1) & (2) & (3) & (4) & (5) & (6) & (7) & (8) \\
X &  Y & pm$_x$ & pm$_y$ & pm$_x$ & pm$_y$ & Proper Motion  & Radial Velocity \\
arcsec & arcsec & (pix) & (pix) & (\kms) & (\kms) & (\kms) & (\kms) \\
\tableline

   25.94     &     -12.02      &        0.4210      &        0.3641     &      33.76     &      29.20      &       47.74     &        -5.50      \\
   24.87     &     -12.82      &        0.3913      &        0.1027     &      31.38     &       8.24      &       44.38     &        -2.80      \\
   24.87     &     -11.92      &        0.2686      &        0.4216     &      21.54     &      33.81      &       30.46     &        -2.80      \\
   24.44     &     -14.12      &        0.4829      &        0.1922     &      38.72     &      15.41      &       54.76     &        -2.80      \\
   23.37     &      12.18      &        0.4522      &       -0.0030     &      36.26     &      -0.24      &       51.28     &        -5.10      \\
   23.05     &     -18.62      &        0.1641      &        0.2169     &      13.16     &      17.39      &       18.61     &         0.20      \\
   22.62     &     -17.42      &        0.1062      &        0.0486     &       8.52     &       3.90      &       12.04     &         0.20      \\
   22.08     &      -2.52      &        0.4363      &       -0.0529     &      34.99     &      -4.24      &       49.48     &        -5.50      \\
   20.91     &      -8.22      &        0.2596      &        0.0559     &      20.82     &       4.48      &       29.44     &         \nodata      \\
   20.59     &     -10.92      &        0.2903      &       -0.1512     &      23.28     &     -12.12      &       32.92     &         \nodata      \\
   20.59     &      -2.72      &        0.4351      &        0.2505     &      34.89     &      20.09      &       49.34     &         \nodata      \\
   20.37     &     -16.82      &       -0.0935      &        0.1247     &      -7.50     &      10.00      &       10.60     &         \nodata      \\
   20.16     &     -17.82      &        0.1391      &        0.1053     &      11.15     &       8.44      &       15.78     &         \nodata      \\
   20.05     &      11.08      &        0.1588      &       -0.1636     &      12.73     &     -13.12      &       18.01     &         \nodata      \\
   20.05     &      11.88      &        0.1974      &       -0.2830     &      15.83     &     -22.69      &       22.39     &         \nodata      \\
   19.94     &      -1.42      &        0.3877      &        0.1159     &      31.09     &       9.29      &       43.97     &         \nodata      \\
   19.94     &      15.08      &        0.2191      &       -0.2212     &      17.57     &     -17.74      &       24.85     &        -3.80      \\
   19.84     &      -7.62      &        0.1277      &        0.2136     &      10.24     &      17.13      &       14.48     &        -5.30      \\
   19.73     &      -9.42      &        0.0458      &        0.2229     &       3.67     &      17.87      &        5.19     &         \nodata      \\
   19.62     &     -10.32      &        0.0679      &        0.0536     &       5.45     &       4.30      &        7.70     &         \nodata      \\

\tableline
\end{tabular}
\end{table*}


\begin{thebibliography}{99}

\bibitem[Artigau et al.(2011)]{Artigau11} Artigau, \'Etienne, Martin, John C., Humphreys, Roberta
  M., Davidson, Kris, Chesneau, Olivier,\& Smith, Nathan. 2011, AJ,
  141, 202A

\bibitem[Baldwin et al.(2000)]{Baldwin00} Baldwin, J. A., Verner, E. M.,
  Verner, D. A., Ferland, G. J., Martin, P. G., Korista, K. T., \&
  Rubin, R. H.  2000, ApJS, 129, 229B

\bibitem[Barker(1978)]{Barker78} Barker, T. 1978, ApJ, 291, 914

\bibitem[Bertin et al.(2002)]{Bertin02} Bertin E., Mellier Y., Radovich M., Missonnier G., Didelon P.,
Morin B., 2002, in Astronomical Data Analysis Software and Systems XI, ASP Conf. Series 281,
228.


\bibitem[Ciardullo et al.(1999)]{Ciardullo99} Ciardullo, R., Bond,
  H. E., Sipior, M. S., Fullton, L. K., Zhang, C.-Y., Schaefer,
  K. G. 1999, \aj, 118, 488

\bibitem[Cahn et al.(1992)]{Cahn92} Cahn, J. H., Kaler, J. B. and
   Stanghellini, L. 1992, AASS. 94, 399

\bibitem[Danehkar et al.(2012)]{Danehkar12} Danehkar, A., Frew, D. J.;
  De Marco, O., \& Parker, Q. A. 2011in "From Interacting Binaries to Exoplanets: Essential
  Modeling Tools", I eds. Richards, M. T. and Hubeny, 2012 IAU Symp. 282, 470


\bibitem[Garc{\'{\i}}a-D{\'{\i}}az et al.(2012)]{GDMT12}
Garc{\'{\i}}a-D{\'{\i}}az, M.~T., L{\'o}pez, J.~A., Steffen,
W., \& Richer, M.~G. 2012, ApJ, 761, 172G

\bibitem[Hajian \& Terzian(1995)]{Hajian95} Hajian, A. R., \& Terzian,
  Y. 1995, AJ, 109, 2600

\bibitem[Heap(1977)]{Heap77} Heap, S. R. 1977, \apj, 215, 864

\bibitem[Kudritzki et al.(1997)]{Kudritzki97} Kudritzki, R. P., Mendez,
  R. H., Puls, J., McCarthy, J. K. 1997, IAUS, 180, 64

\bibitem[Li et al.(2002)]{Li02} Li J., Harrington J.P. \& Borkowski K.J. 2002, AJ, 123, 2676L

\bibitem[Liller \& Liller(1968)]{Liller68}
Liller, M. H., \& Liller, W. 1968, in IAU Symp. 34, Planetary Nebulae, ed. D. E.
Osterbrock \& C. R. O’Dell (Dordrecht: Reidel), 38 (LL68)

\bibitem[L\'opez et al.(2012)]{Lopez12} L\'opez, J. A., Richer, M. G.,
  Garc\'ia-D\'iaz, M. T., Clark, D. M., Meaburn, J., Riesgo, H., Steffen,
  W., \& Lloyd, M. 2012, RevMexAA, 48, 3

\bibitem[Maciel(1981)]{Maciel81} Maciel, W. J.\ 1981 Astro. Ap. Suppl., 44, 123
Gon{\c c}alves, D.~R., Villaver, E., Mampaso, A., Perinotto, M., Schwarz,
H. E., \& Zanin, C. 2000, \apj, 535, 823

\bibitem[Guerrero et al.(2005)]{Guerrero05}
Guerrero, M.A., Chu, Y.-H., Gruendl, R.A., Meixner, M., 2005, \aap, 430, L69-L72

\bibitem[Meaburn et al.(2003)]{Meaburn03} Meaburn, J., L\'opez,
J. A., Guti\'errez, L., Quir\'oz, F., Murillo, J. M. \& Vald\'ez, J.
2003, RMxAA, 39, 185M

\bibitem[M\'endez et al.(2011)]{Mendez11} M\'endez, R. H., Urbaneja,
  M.A., Kudritzki, R. P. \& Prinja, R. K. 2011, in IAU Symp 283, Planetary Nebula: An Eye to the Future, ed. A. Manchado, L. Stanghellini, \& D. Schoenberner (Cambridge University Press), 436

\bibitem[Natta et al.(1980)]{Natta80} Natta, A., Pottasch, S. R., \&
  Preite-Martinez, A., 1980, \aap, 84, 284

\bibitem[O'Dell et al.(2013)]{Odell13} O'Dell, C. R., Ferland, G. J., Henney, W. J., \&
  Peimbert, M. 2013, AJ, 145, 170

\bibitem[O'Dell et al.(2008)]{odell08} O'Dell, C. R. \& Henney, W. J.,
  2008, AJ, 136, 1566

\bibitem[O'Dell et al.(1990)]{ODell90} O'Dell, C. R., Weiner,
  L. D., \& Chu, Y. H. 1990, ApJ, 362, 226

\bibitem[Pauldrach et al.(2004)]{Pauldrach04} Pauldrach, A. W. A.,
  Hoffmann, T. L., \& Mendez, R. H. 2004, \aap, 419, 1111


\bibitem[Pottasch et al.(2011)]{Pottasch11} Pottasch, S. R., Surendiranath, R. \& Bernard-Salas, J.
2011, A\&A, 23, 531A

\bibitem[Pottasch et al.(2008)]{Pottasch08} Pottasch, S. R.,  Bernard-Salas, J.  \& Roelling T. L.  2008, \aap, 481, 393

\bibitem[Ruiz et al.(2013)]{Ruiz12}
Ruiz, N., Chu, Y.-H., Gruendl, R.A., Guerrero, M.A., Jacob, R., Schönberner, D., Steffen, M., 2013, \apj, 767, 35, 11pp.

\bibitem[Tinkler \& Lamers(2002)]{Tinkler02} Tinkler, C. M., Lamers,
  H. J. G. L. M. 2002, \aap, 384, 987

\bibitem[Ueta et al.(2006)]{Ueta06} Ueta, Toshiya, Murakawa, Koji, Meixner, Margaret. 2006,
  ApJ, 641, 1113U

\bibitem[Stanghellini et al.(2008)]{Stanghellini08}
  Stanghellini, L., Shaw, R. A., \& Villaver, E. 2008, ApJ, 689, 194

\bibitem[Steffen et al.(2011)]{Steffen11} Steffen, Wolfgang, \& Koning, Nico. 2011, AJ, 141, 76S

\bibitem[Szyszka et al.(2011)]{Szy11} Szyszka, C., Zijlstra, A. A., \&
  Walsh, J. R. 2011, MNRAS, 416, 715S





\end{thebibliography}
\end{document}